\documentclass[aps,reprint,amsmath,amssymb,pra,twocolumn,superscriptaddress]{revtex4-2}
\usepackage{graphicx}
\usepackage{dcolumn}
\usepackage{bm}
\usepackage{float}
\usepackage{xcolor}
\usepackage{lipsum}
\usepackage{comment}
\usepackage{lineno}

\newcommand{\p}{\partial}

\newcommand{\ep}{\varepsilon}
\newcommand{\vta}{\vartheta}
\newcommand{\om}{\omega}

\newcommand{\vD}{\varDelta}
\newcommand{\nn}{\nonumber}
\newcommand{\ta}{\theta}

\newcommand{\wt}{\widetilde}
\newcommand{\wh}{\widehat}

\newcommand{\cH}{{\cal H}}
\newcommand{\cF}{{\cal F}}

\newcommand{\cW}{{\cal W}}

\newcommand{\be}{\begin{equation}}                                       
	\newcommand{\ee}{\end{equation}}
\newcommand{\ba}{\begin{eqnarray}}
	\newcommand{\ea}{\end{eqnarray}}
\newcommand{\bref}[1]{(\ref{#1})}

\newcommand{\bi}[1]{\bibitem{#1}}\newcommand{\lab}[1]{\label{#1}}

\newcommand{\bsub}{\begin{linenomath}\begin{subequations}}                      
		\newcommand{\esub}{\end{subequations}\end{linenomath}}

\begin{document}
	
	\preprint{APS/123-QED}
	\title{
Parametric conversion  
via second harmonic generation and two-hump solitons
in phase-matched microresonators}

	\author{Vladislav V. Pankratov}
\affiliation{Department of Physics, University of Bath, Bath, BA2 7AY, UK} 
\affiliation{Centre for Photonics and Photonic Materials, University of Bath, Bath, BA2 7AY, UK}
	\author{Dmitry V. Skryabin}
	\email{d.v.skryabin@bath.ac.uk}
	\affiliation{Department of Physics, University of Bath, Bath, BA2 7AY, UK} 
	\affiliation{Centre for Photonics and Photonic Materials, University of Bath, Bath, BA2 7AY, UK}

	\begin{abstract}
Frequency conversion in microresonators has revolutionised modern-day nonlinear and quantum optics. Here, we present a theory of the multimode second harmonic generation in microresonators under conditions when the parametric conversion back to the pump spectrum dominates through the large domain in the resonator parameter space. We demonstrate that the spectral tunability of the sideband generation in this regime is governed by a discrete sequence of the so-called Eckhaus instabilities. We report the transition to modelocking and generation of solitons, which have a double-pulse structure in the pump field. These solitons exist outside the bistability interval and on a slightly curved  background.
	\end{abstract}
	\maketitle
		
\section{Introduction}
Frequency conversion and comb generation in optical resonators and microresonators are transforming research and applications of nonlinear and quantum optics. They impact precision frequency metrology, modelocking, soliton photonics, quantum and classical processing of information~\cite{rev1,qc,qc1}. Microresonators made with the materials possessing the second-order, i.e., $\chi^{(2)}$, nonlinearity allow generating frequency combs at twice or half of the pump frequency at the comparatively low input powers, see, e.g.,~\cite{mark,wab,hong1,hong2,ingo1,ingo2}. Another advantage of $\chi^{(2)}$ resonators is that the side-band generation relying on the $\chi^{(2)}$ effects does not critically depend 
on the dispersion sign, which facilitates working with visible and near-infrared sources. 
Though the generation of $\chi^{(2)}$ solitons in micro- and  bow-tie resonators has been demonstrated~\cite{hong2,marandi}, it remains far from being as developed as the soliton techniques in Kerr resonators~\cite{rev2}. This, in its turn, hinders the development of applications of the $\chi^{(2)}$ based soliton modelocking in microresonators.

The initiation of the Kerr frequency comb happens via the generation of two weak sidebands $\om_{\pm\mu}$ from the two pump photons $\om_p$, i.e., $\hbar\om_p+\hbar\om_p=\hbar\om_\mu+\hbar\om_{-\mu}$, while the onset of the $\chi^{(2)}$ mediated side-band generation is more complex. Our recent analysis in Ref.~\cite{pra} shows that the operator driving the initial stage of the second harmonic generation in the multimode $\chi^{(2)}$ resonators involves simultaneous and interdependent sum-frequency and parametric down-conversion terms. If the pump frequency is  $\om_p$, then its second harmonic, $2\om_p$,  is allowed to down-convert via the parametric process to the two sidebands of the pump field, i.e., 
\be
\hbar 2\om_p=\hbar\om_{\mu a}+\hbar\om_{-\mu a}.
\lab{e1}
\ee
Here and below, the  subscripts 'a' and 'b' are used for the pump and second harmonic sidebands, respectively. $\mu=0,\pm 1,\pm 2,\dots$ numbers the sideband orders. The laser photon can also sum up with its own sideband  to generate sidebands of the second harmonic, i.e.,
\be
\hbar\om_p+\hbar\om_{\mu a}=\hbar\om_{\mu b}.
\lab{e2}
\ee

The magnitude of the $\om_{\mu a}$ to $\om_{-\mu a}$ 
sideband coupling, controlling the rate 
of the parametric down-conversion in Eq.~\bref{e1}, 
is determined by the amplitude of the second harmonic, which
is critically susceptible to the phase-matching.
On the other hand, the magnitude of the $\om_{\mu a}$ to $\om_{\mu b}$ 
coupling, controlling the rate of the sum-frequency process 
in Eq.~\bref{e2}, is determined by the pump amplitude~\cite{pra}. Thus, by tuning the phase-matching, one can adjust the balance between parametric and  sum-frequency processes. 

Ref.~\cite{pra} considers the case when the sum-frequency dominates the parametric conversion. It features the Rabi oscillations between the $\om_{\mu a}$ and $\om_{\mu b}$ states assisting with the quantum memory designs~\cite{rabi1,rabi2} and photon-photon polariton quasiparticles~\cite{prr}. It also shows that the frequency comb generation thresholds and soliton existence conditions
can be derived by applying the formalism of the dressed states 
similar to the two-level atom theory~\cite{pra}.
Present work is a companion to Ref.~\cite{pra}, where a comprehensive list of modern and historical references on $\chi^{(2)}$ effects in resonators can be found. 

Here, we deal with the resonator geometry as in Ref.~\cite{pra} but consider the case of near phase-matching, such that the sum-frequency and parametric conversion rates become comparable. However, both rates are typically inferior to the much faster frequency scale set by the difference in the repetition rates of the first and second harmonics. This resonator operation condition is common in practice, see, e.g., Refs.~\cite{apl,hong3}, however, the theory for the multi-mode second harmonic generation under such conditions has remained unknown and is developed below. 

The resonator considered here has $\sim 20$GHz repetition rate and  $\sim 1$GHz repetition rate difference of the first and second harmonics, which implies that the initially overlapping $\om_p$ and $2\om_p$ pulses would be on the opposite sides of the linear resonator only after a few tens of round trips. Since the repetition rate difference characterizes how fast the pulses walk away one from the other, it is called the walk-off parameter. The 'walk-off' term is common and is also shared between the fibre and other resonator contexts, see, e.g.~\cite{walk1,walk2,walk3,walk4}.

The methodology applied below is inherited from our recent theory of the parametric processes in the microresonator half-harmonic generation~\cite{comphys}. This approach relies on the slowly (relative to $1$GHz) varying amplitudes leading to the analytically solvable coupled-mode model. The two-colour bright soliton pulses are hard to expect for such a large walk-off. However, they exist and have a two-hump structure in the pump field.

\section{Model}
We assume that the pump laser, $\om_p$, is tuned around 
the resonator mode with the frequency $\om_{0a}$, so that 
the multimode intra-resonator pump field
and its second-harmonic are expressed as 
\be
\begin{split}
	&A e^{iM\vta-i\om_p t}+c.c.,~A=\sum_\mu a_\mu(t) e^{i\mu\ta},\\ 
	&B e^{i2M\vta-i2\om_p t}+c.c.,~B=\sum_\mu b_\mu(t) e^{i\mu\ta},\\&\ta=\vta- D_{1}t.
\end{split}
\lab{field}
\ee
Here, $M$ is the absolute mode number corresponding to the frequency 
$\om_{0a}$, $\vta=(0,2\pi]$ is the angular coordinate, 
$\ta$ is the coordinate in the reference frame rotating with the user defined rate $D_1$, and 
$\mu=0,\pm 1,\pm 2,\dots$ is the relative mode number. $a_\mu$, $b_\mu$ 
are the amplitudes of the pump (fundamental) and second-harmonic modes. 
The respective resonator frequencies are
\be
\om_{\mu \zeta}=\om_{0\zeta}+\mu D_{1\zeta}+\tfrac{1}{2}\mu^2 D_{2\zeta},~\zeta=a,b,
\lab{om}\ee
where,  $D_{1\zeta}$ are the linear repetition rates  
and $D_{2\zeta}$ are dispersions. 
In what follows, we choose  $D_1=D_{1a}$. 
$D_{1b}-D_{1a}$ is the walk-off parameter, 
i.e., the repetition rate difference.

Coupled-mode equations governing the evolution of $a_\mu(t)$, $b_\mu(t)$ are~\cite{josab}
\be
\begin{split}
	i\p_t a_{\mu}=\delta_{\mu a}a_{\mu} -& \frac{i\kappa_a}{2}
	\big(a_{\mu}-\wh\delta_{\mu,0}\cH\big)\\ 
	& -\gamma_a\sum_{\mu_1 \mu_2}\wh\delta_{\mu,\mu_1-\mu_2}b_{\mu_1}a^*_{\mu_2},
	\\
	i\p_tb_{\mu}=\delta_{\mu b}b_{\mu} - &\frac{i\kappa_b}{2}
	b_{\mu}
	-\gamma_b\sum_{\mu_1 \mu_2}\wh\delta_{\mu,\mu_1+\mu_2}a_{\mu_1}a_{\mu_2},
\end{split}
\lab{tp1}
\ee
where $\wh\delta_{\mu,\mu'}=1$  for $\mu=\mu'$ and is zero otherwise.
$\cH$ is the pump parameter under the critical coupling conditions,
$\cH^2=\cF\cW/2\pi$, where
$\cW$ is the laser power, and $\cF=D_{1a}/\kappa_a$ is finesse.  
$\delta_{\mu\zeta}$ are the modal detuning parameters in the rotating reference frame,
$\delta_{\mu a}  =(\om_{\mu a}-\tfrac{1}{2}\om_p)-\mu D_{1a}$ and $
\delta_{\mu b}  =(\om_{\mu b}-\om_p)-\mu D_{1a}$.
 $\delta_{\mu \zeta}$  can be expressed via  pump detuning, $\delta_{0a}$, 
and the frequency-matching, i.e., phase-matching parameter, $\ep$,~\cite{pra}
\be
\begin{split}
	\ep&=2\om_{0 a}-\om_{0b},\\
	\delta_{\mu a}&=\om_{0a}-\om_p+\tfrac{1}{2}\mu^2D_{2a}\\ &=\delta_{0a}+
	\tfrac{1}{2}\mu^2D_{2a},\\
	\delta_{\mu b}  &=
	\om_{0 b}-2\om_p+\mu( D_{1b}-D_{1a})+\tfrac{1}{2}\mu^2 D_{2b}\\
	&=
	2\delta_{0a}-\ep+\mu( D_{1b}-D_{1a})+\tfrac{1}{2}\mu^2 D_{2b}.
\end{split}
\lab{fi1}
\ee

We assume that the microresonator is made to operate close to phase-matching,  $\ep=0$, while the repetition rate difference, $D_{1b}-D_{1a}$, is the dominant frequency scale  in Eq.~\bref{fi1}, i.e.,
\be
\mu|D_{1a}-D_{1b}|\gg |\ep|,~\mu^2|D_{2\zeta}|,~\kappa_\zeta.
\lab{e09}
\ee
The parameter values used by us
are typical for the LiNbO$_3$ 
whispering gallery microresonators~\cite{ingo1,apl} and are listed in Table~\ref{t1}, so that one can verify the validity of Eq.~\bref{e09}.

\begin{table}[t]
		\begin{tabular}{l}
			Linewidth: $\kappa_a/2\pi=1$MHz, $\kappa_b/2\pi=4$MHz 
			\\ \hline 
			Pump finesse: $D_{1a}/\kappa_a=20000$
			\\ \hline 
			Repetition rate difference: $(D_{1a}-D_{1b})/2\pi=1$GHz
			\\ \hline 
			Dispersion: $D_{2a}/2\pi=-100$kHz, $D_{2b}/2\pi=-200$kHz
			\\ \hline 
			Nonlinearity: $\gamma_{a}/2\pi=\gamma_{b}/2\pi=300\text{MHz}/\sqrt{\text{W}}$
			\\ \hline
			Phase-mismatch: $\ep/2\pi$ between $-50$ and $+50~$MHz
			\\ \hline
			Scaling of intraresonator power: 
			$\cH_*^2=\kappa_a\kappa_b/\gamma_a\gamma_b=44\mu$W
			\\ \hline
			Laser power: $\cW$ between $\sim 1\mu\text{W}$ and $\sim 1$mW
		\end{tabular}
	\caption{Resonator and laser parameters used in this work. The values are representative for a whispering gallery LiNbO$_3$ microresonator pumped at $1\mu$m.
	}
	\lab{t1}
\end{table}

We now make the substitution, 
\be
b_{\mu}(t)= B_{\mu}(t)e^{-i\mu(D_{1 b}-D_{1 a})t},~ \mu\ne 0,
\lab{bb}
\ee
where $B_{\mu}(t)$ are the slowly varying amplitudes relative to the fast oscillating exponent, and
apply the slowly varying approximation by eliminating all the 
terms oscillating with the high frequencies, see mathematical details in Ref.~\cite{comphys}. This procedure 
yields the following system of equations
	\bsub
	\lab{cm}
\begin{align}
&i\p_t a_{0 }=\kappa_a\vD_{0  a}a_{0 }-\gamma_ab_{0 }a^*_{0}+ \frac{i\kappa_a}{2}
\cH,
\lab{p91}	\\	
&i\p_t a_{\mu }=\kappa_a\vD_{\mu  a}a_{\mu } 	
-\gamma_ab_{0 }a^*_{-\mu },~\mu\ne 0,
\lab{p92}	\\
&i\p_t a_{-\mu }=\kappa_a\vD_{\mu  a}a_{-\mu} 	
-\gamma_a b_{0 }a^*_{\mu },
\lab{p93}	\\
&i\p_t b_{0 }=\kappa_b\vD_{0 b}b_{0 }-\gamma_ba_{0 }^2 -	2\gamma_b\sum_{\mu> 0}a_{\mu}a_{-\mu}.
\lab{p94}
\end{align}
\esub
Fig.~\ref{f0} illustrates the dominant processes well described by Eq.~\bref{cm}. 
The second harmonic sidebands,  $b_{\mu\ne 0}$, are not featured in Eq.~\bref{cm} and, start playing a role either in the higher-order approximations or if the assumptions in Eq.~\bref{e09} are violated. The reduced model, Eq.~\bref{cm}, is applied below to derive the analytical results, while all the numerical data have been generated using the master model, Eq.~\bref{tp1}.

The linear spectrum in Fig.~\ref{f0} should be contrasted with the dressed, i.e.,  nonlinearity modified, spectrum implicated in the interplay of the Rabi oscillations and parametric processes studied in Ref.~\cite{pra} and illustrated in Fig.~6 there.
Simultaneous parametric conversion and second harmonic generation and  other multistep $\chi^{(2)}$ frequency conversion schemes have long  history outside the comb context and continue to be explored  now days~\cite{old,new1,new2,new3}.

Eq.~\bref{cm} and text below use the $\vD_{\mu\zeta}$ parameters, which are the auxiliary dimensionless detunings,
\be
\vD_{\mu\zeta}=\left(
\delta_{0 \zeta}+\frac{1}{2}\mu^2D_{2\zeta}-i\frac{1}{2}\kappa_\zeta\right)
\frac{1}{\kappa_\zeta}.
\lab{fiel}
\ee
Eq.~\bref{fiel} includes the losses and, hence, $\vD_{\mu\zeta}$ are complex-valued.
We note that $\vD_{\mu\zeta}$ are free from the walk-off parameter, $D_{1b}-D_{1a}$, 
absorbed by the fast oscillating exponents in Eq.~\bref{bb}.
The sum-frequency rate is determined by $\gamma_b |a_0|$, see Eq.~\bref{p94}, 
and the rate of the parametric gain is $\gamma_a |b_0|$, 
see Eqs.~\bref{p91}-\bref{p93}, which are also assumed to comply with
\be
 \mu|D_{1a}-D_{1b}|\gg \gamma_a|b_0|,~\gamma_b|a_0|.
	\lab{e09a}
\ee

The difference between Eqs.~\bref{p91}-\bref{p94} and the directly pumped OPO model considered in Ref.~\cite{comphys} is in the moving the pumping term between the $a_0$ and $b_0$ equations. However,  this change leads to a different and more involved  theory than in Ref.~\cite{comphys}. 
On the other hand, the currently applied $\mu|D_{1a}-D_{1b}|\gg |\ep|$ condition, see Eq.~\bref{e09}, simplifies the matters relative to the second-harmonic theory  for $\mu|D_{1a}-D_{1b}|\sim |\ep|$ developed in Ref.~\cite{pra}.

\begin{figure}[t]
	\centering
	\includegraphics[width=0.45\textwidth]{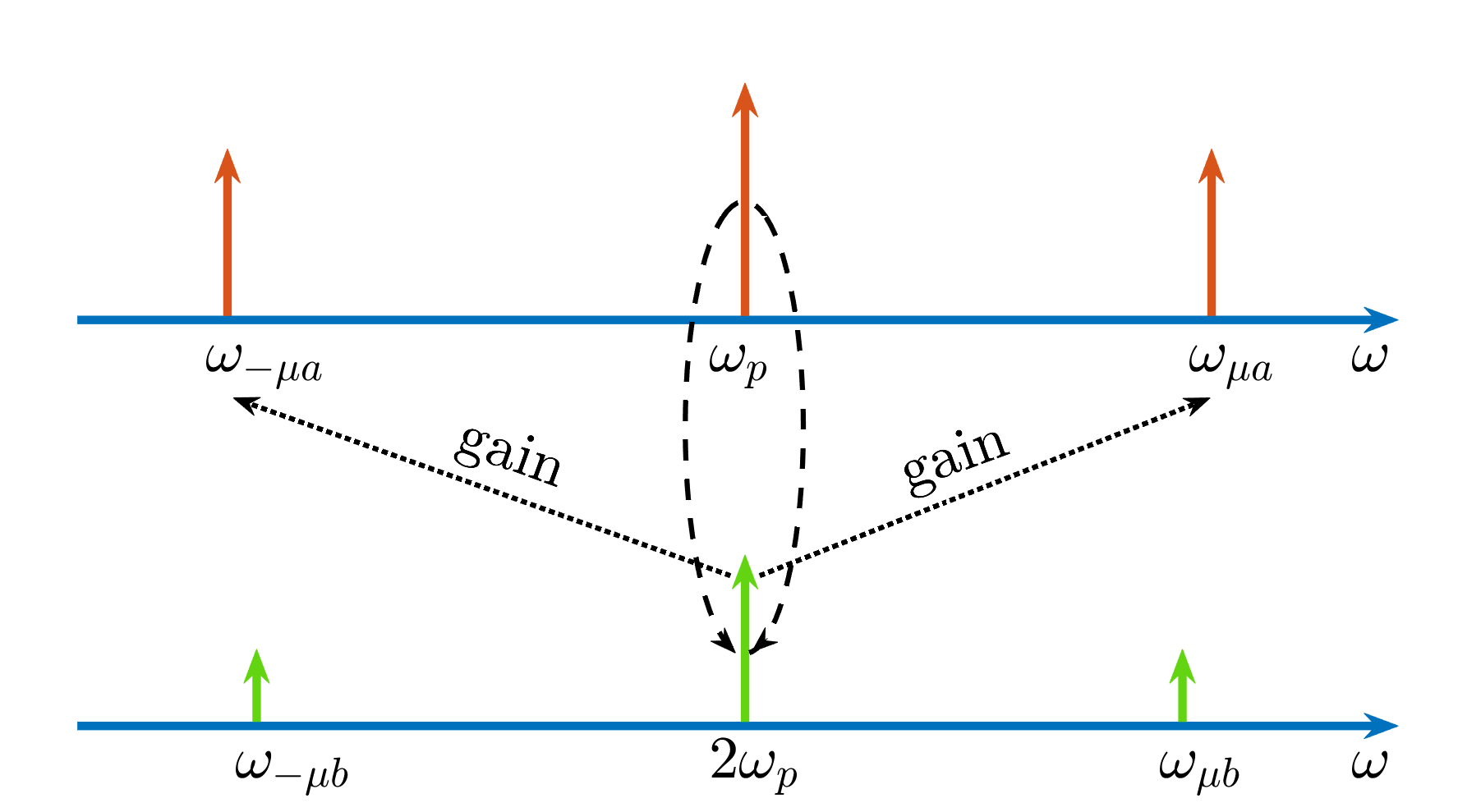}	     
	\caption{Illustration of one of 
		parametric processes captured by the reduced model valid for 
		$\mu |D_{1a}-D_{1b}|\gg |\ep|$, see Eq.~\bref{cm}. It should be compared 
		with the dressed state theory for 
		$\mu |D_{1a}-D_{1b}|\sim |\ep|$ illustrated by Fig.~6 in Ref.~\cite{pra}. 
		Here, $\ep$ is the phase-matching parameter. }
	\lab{f0}
\end{figure}

 \begin{figure*}[t]
	\centering
	\includegraphics[width=1.\textwidth]{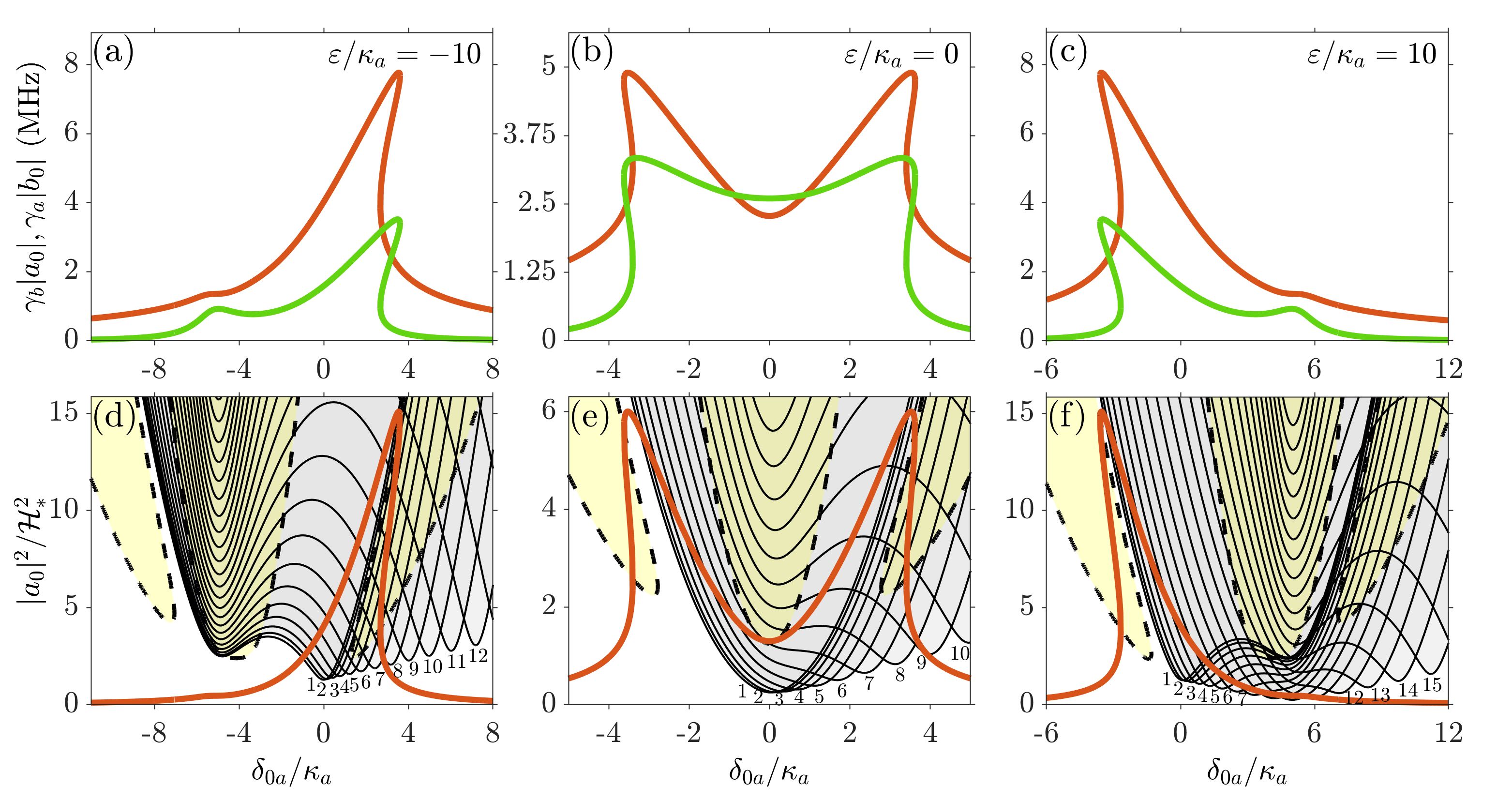}	     
	\caption{(a-c) Amplitudes of the homogeneous (single mode) states  vs detuning for the negative, zero and positive values of the phase-mismatch parameter, $\ep$. The red (dark gray) lines mark $\gamma_b|\wt a_0|$, which is the rate of the sum-frequency process. The green (light gray) lines mark $\gamma_a|\wt b_0|$, which is the rate of the parametric process. Laser power is $\cW=0.7\mu$W, which corresponds to  $\cH^2/\cH_*^2=50$. (d-f): Instability boundaries of the homogeneous state. The black lines, $|a_0|^2=|\hat a_{0,\text{th}}^{(\mu)}|^2$, and grey areas show the $\mu\ne 0$ Benjamin-Feir instabilities. The dashed black lines and yellow (light gray) shading show the $\mu=0$ instabilities.  Homogeneous state becomes unstable, when the red (gray) line, $|a_0|^2=|\wt a_0|^2$, crosses any of the thresholds.}
	\lab{f1}
\end{figure*}

\section{Theory of parametric states}

\subsection{Homogeneous state}
From the diversity of possible solutions 
of Eqs.~\bref{p91}-\bref{p94} the simplest one
is the  spatially homogeneous state with 
\be A=\wt a_{0},~B=\wt b_{0}.\ee
We added the tildes here to introduce the notational difference 
with the  $\mu=0$ modes entering the OPO
states introduced below and marked with the hats. 

Setting the sideband amplitudes to zero, $a_\mu=0$, the solution of
Eq.~\bref{p94} is
\be
\wt b_0=\frac{\gamma_b\wt a_0^2}{\kappa_b\vD_{0b}}.
\ee 
Then, Eq.~\bref{p91} can be rearranged as
\be
\left(\vD_{0a}-\frac{|\wt a_{0}|^2}
{\vD_{0b}\cH_*^2} \right) \frac{\wt a_{0}}{\cH_*}=
\frac{-i\cH}{2\cH_*}.
\lab{qn}
\ee
Here,
\be
\cH_*^2=\dfrac{\kappa_a\kappa_b}{\gamma_a\gamma_b}=44\mu\text{W}~\lab{nor}
\ee
is the characteristic power introduced for the normalisation purposes.
Taking modulus squared of Eq.~\bref{qn} we find the real cubic 
equation for $|\wt a_0|^2$,
\ba
\frac{|\wt a_0|^6}{\cH_*^6}-2\text{Re} \left(\vD_{0 a} \vD_{0 b} \right)
\frac{|\wt a_0|^4}{\cH_*^4}
+&&|\vD_{0 a} \vD_{0 b}|^2\frac{|\wt a_0|^2}{\cH_*^2}
\nn\\
=&&\frac{\cH^2|\vD_{0 b}|^2}{4\cH_*^2}.\lab{cub}
\ea

Examples of the dependencies of $\gamma_b|\wt a_0|$ (red lines) and $\gamma_a|\wt b_0|$ (green lines) vs the pump detuning $\delta_{0a}$ are shown in Figs.~\ref{f1}(a)-\ref{f1}(c). One can see that sum-frequency (red) and parametric (green) rates are of the same order. This should be compared with Fig.~1 in Ref.~\cite{pra}, where the phase-matching parameter was tuned to $|\ep|\sim\mu|D_{1a}-D_{1a}|$ to achieve the sum-frequency rates dominating the parametric ones by two orders of magnitude.

\subsection{Benjamin-Feir instabilities and emergence of OPO (Turing pattern) states}
OPO states correspond to the resonator generating one, or primarily one, 
sideband pairs in the pump field, i.e., 
\be A= 
\hat a_0+a_{\nu}e^{i\nu\ta}+a_{-\nu}e^{-i\nu\ta},~B= \hat b_{0}.
\lab{op}
\ee
The OPO states are thus the simplest representatives of the Turing patterns, which are expected to bifurcate from the homogeneous state.
Eqs.  \bref{p92}-\bref{p94} are resolved as
\ba
&& a_{\nu}=|a_{\nu}|e^{i\phi_\nu},~
a_{-\nu }=|a_{\nu }|,~
\hat b_0=\frac{\vD_{\nu a}\kappa_a}{\gamma_a}e^{i\phi_\nu},
\lab{t2b}\\
&& e^{i\phi_\nu}
=\frac{\hat a_{0}^2/\cH_*^2}{\vD_{0b}\vD_{\nu a}-2|a_\nu|^2/\cH_*^2}.
\lab{t33}
\ea
We note that the homogeneous state is the same for the master, Eq.~\bref{tp1}, and  reduced, Eq.~\bref{cm},  models, but the above OPO states are exact 
for the reduced and approximate the master model.

\begin{figure}[t]
	\centering
	\includegraphics[width=0.48\textwidth]{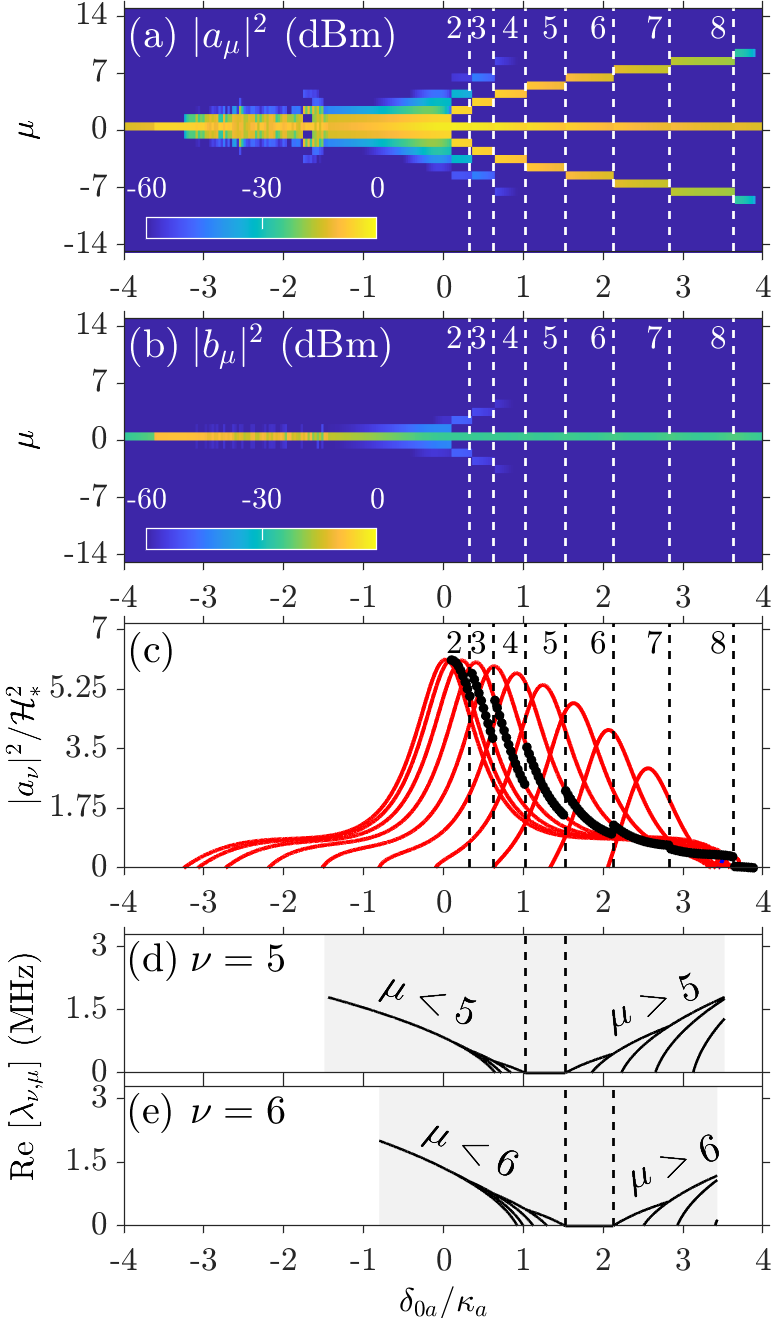}	     
	\caption{(a,b) Results of dynamical simulations of Eq.~\bref{tp1}. Panel (a) shows 
		the mode number spectrum for the pump and panel (b) for the second harmonic  field, $\cH^2/\cH_*^2=50$ ($\cW\approx 0.7\mu$W), $\ep=0$. (c) Side-band amplitudes of the pump field found using Eq.~\bref{x3a} (red (gray) lines) and from the data in the panel (a) (black circles). (d,e) Eckhaus instability growth rates, $\lambda_{\nu,\mu}$, for the $\nu=5$ and $\nu=6$ OPO states. The OPO states exist in the shaded area and stable between  the dashed  lines.
	}
	\lab{f2}
\end{figure}

The modulus squared of Eq.~\bref{t33} yields the 
quadratic equation for the sideband amplitudes
expressing $|a_\nu|^2$ as a function of $|\hat a_0|^2$,
\ba
 \frac{|a_{\nu}|^2}{\cH_*^2}	=&&\frac{\text{Re}(\vD_{\nu a}\vD_{0 b})}{2}
 \lab{x3a}\\
+&&\frac{(-1)^j}{2}\sqrt{|\hat a_0|^4/\cH_*^4
	-\text{Im}^2(\vD_{\nu a}\vD_{0 b})},~j=0,1,
\nn\ea
where the value of $|\hat a_0|^2$ remains unknown and will be determined in the next subsection.

Setting  sideband powers to zero,  $|a_{\nu}|^2=0$  in Eq.~\bref{x3a}, 
we find 
\be
\frac{|\hat a_{0,\text{th}}^{(\nu)}|^2}{\cH_*^2}=|\vD_{0 b} \vD_{\nu a}|,
\lab{eq7}\ee
which gives a  
sequence of $|\hat a_{0,\text{th}}^{(\nu)}|^2$ vs $\delta_{0a}$ lines 
for all possible $\nu\ne 0$ corresponding to the 
thresholds where the OPO states (Turing patterns) bifurcate 
from the no-OPO (homogeneous) state. Thus, condition 
\be
|\wt a_0|^2=|\hat a_{0,\text{th}}^{(\nu)}|^2
\lab{e77}
\ee
gives the mode number specific Benjamin-Feir bifurcation thresholds 
of the homogeneous state.
Figs.~\ref{f1}(d)-\ref{f1}(f) show how the 
$|a_0|^2=|\wt a_0|^2$ vs $\delta_{0a}$ line intersects with a sequence of 
the $|a_0|^2=|\hat a_{0,\text{th}}^{(\nu)}|^2$  lines. 
Each intersection point of the red and grey lines corresponds to the threshold detuning for generating the corresponding side-band pair. 

We recall that the theory of this section is applicable for $\mu\ne 0$.
To find the $\mu=0$ instability boundary and verify the theory's results for $\mu\ne 0$, we have linearised the master model, Eq.~\bref{tp1}, around the homogeneous state.  To achieve this we have assumed~\cite{pra} 
\be
\begin{split}
& 
A=\wt a_0
+x_{\mu a} e^{\lambda_\mu  t+i\mu\ta}
+y_{\mu a}^*e^{\lambda_\mu t-i\mu\ta},\\ 
& 
B=\wt b_0
+x_{\mu b}e^{\lambda_\mu  t+i\mu\ta}
+y_{\mu b}^*e^{\lambda_\mu t-i\mu\ta}.
\end{split}
\lab{bf}
\ee 
Here, $\vec x_\mu=(x_{\mu a},x_{\mu b},y_{\mu a},y_{\mu b})^T$ is the perturbation vector and
Re$\lambda_\mu$ is the instability growth rate.
The resulting eigenvalue problem for  $\vec x_\mu$ is 
\ba
&&i\lambda_\mu\vec x_\mu=\lab{ev}\\
&&\left[\begin{array}{cccc}
	\delta_{\mu a}-\tfrac{i\kappa_a}{2}&-\gamma_a \wt a_0^*&-\gamma_a\wt b_0&0\\
	-2\gamma_b \wt a_0&\delta_{\mu b}-\tfrac{i\kappa_b}{2}&0&0\\
	\gamma_a\wt b_0^*&0&-\delta_{-\mu a}-\tfrac{i\kappa_a}{2}&\gamma_a \wt a_0\\
	0&0&2\gamma_b \wt a_0^*&-\delta_{-\mu b}-\tfrac{i\kappa_b}{2}
\end{array}\right]\vec x_\mu.
\nn
\ea

The above matrix makes it obvious that $\wt b_0$ couples the 'a'-sidebands and drives the parametric conversion, see Eq.~\bref{e1},  and $\wt a_0$ provides the  'a' to 'b' coupling and drives the sum-frequency process, see Eq.~\bref{e2}. The Re$\lambda_0=0$ instability boundary is shown in Figs.~\ref{f1}(d)-(f) with the dashed-black lines and yellow shading. The Re$\lambda_{\mu\ne 0}=0$ Benjamin-Feir instability boundaries found from  Eq.~\bref{ev} and  derived from Eq.~\bref{e77} can not be distinguished; see the full-black lines and grey shading in Figs.~\ref{f1}(d)-(f). 

We recall that the $\delta_{\mu\zeta}$ detunings in Eq.~\bref{ev} include the walk-off parameter, see Eq.~\bref{fi1}. 
$\delta_{\mu\zeta}$ should be compared with the walk-free $\vD_{\mu\zeta}$, see  Eq.~\bref{fiel}, used to get our analytical results. Thus, we are dealing with a case where the large walk-off does not play a role in destabilising the homogeneous state, cf., ~\cite{walk2}. The independence of the instabilities from walk-off agrees with the fact that the second harmonic field remains quasi-monochromatic, see Fig.~\ref{f2}(b), and hence should not be subjected to the group-velocity related effects.

Figures~\ref{f2}(a) and \ref{f2}(b) show the data from the numerical simulation of the master model, Eq. \bref{tp1}, for the fixed input power and varying detuning. Comparing the pump and second harmonic spectra confirms our analytical result that the second harmonic sidebands are tiny and the ones in the pump are much stronger. This illustrates the essence of what we term the second-harmonic driven parametric conversion. The transition to a sequence of the OPO states for $\delta_{0a}>0$ is evident from the data in Fig.~\ref{f2}(a). One could plausibly assume that, as in a directly pumped OPO~\cite{comphys}, the switching between the different OPO states is governed by the Eckhaus instabilities of the OPO states and not by the Benjamin-Feir bifurcations of the homogeneous state, see  Section~IV for details.

\subsection{ Super- and sub-critical bifurcations}
The OPO (Turing pattern) states can emerge either super- or sub-critically at their bifurcation points from the homogeneous state. Substituting $\hat b_0$ and $e^{i\phi_\nu}$ from Eqs.~\bref{t2b}, \bref{t33} 
in Eq.~\bref{p91} and rearranging, we find
\be
\left(\vD_{0a}-\dfrac{\vD_{\nu a}|\hat a_{0}|^2/\cH_*^2}
{\vD_{\nu a}\vD_{0b}-2|a_\nu|^2/\cH_*^2} \right) \frac{\hat a_{0}}{\cH_*}=
\frac{-i\cH}{2\cH_*}.
\lab{qe}
\ee
Comparing the above and Eq.~\bref{qn}, 
we immediately see that they predictably coincide 
for $|a_\nu|^2=0$, i.e., 
$\hat a_0=\wt a_0$ at the bifurcation points. Taking modulus squared
of Eq.~\bref{qe} and using Eq.~\bref{x3a} for $|a_\nu|^2$ gives an explicit 
dependence of $\cH^2/\cH_*^2$ vs $|\hat a_0|^2$ for the Turing pattern states,  
see Fig.~\ref{f3}(a) for the $\nu=10$ case. 
Substituting the $|\hat a_0|^2$ vs  $\cH^2/\cH_*^2$ data set 
in Eq.~\bref{x3a} makes up the corresponding sideband 
powers,  see Fig.~\ref{f3}(b). The transition between the full and dotted red lines in Fig.~\ref{f3} shows how the $j=0$ and $j=1$ realisations of Eq.~\bref{x3a} play a role 
in constructing the complete branch of the Turing pattern. 

The fact that the Turing pattern branch folds at $\cH^2=\cH_{\nu,\text{fold}}^2$,
see Fig.~\ref{f3}(b), points that the patterns can bifurcate from the homogeneous state either super- (no fold) or sub-critically. In the latter case, a range exists in the parameter space where a given sideband $\nu$ can have two different powers, with the lower one never being stable. 
Quite remarkably, the fold condition and the Turing pattern solutions can be expressed explicitly via the resonator parameters. 

\begin{figure}[t]
	\centering
	\includegraphics[width=0.49\textwidth]{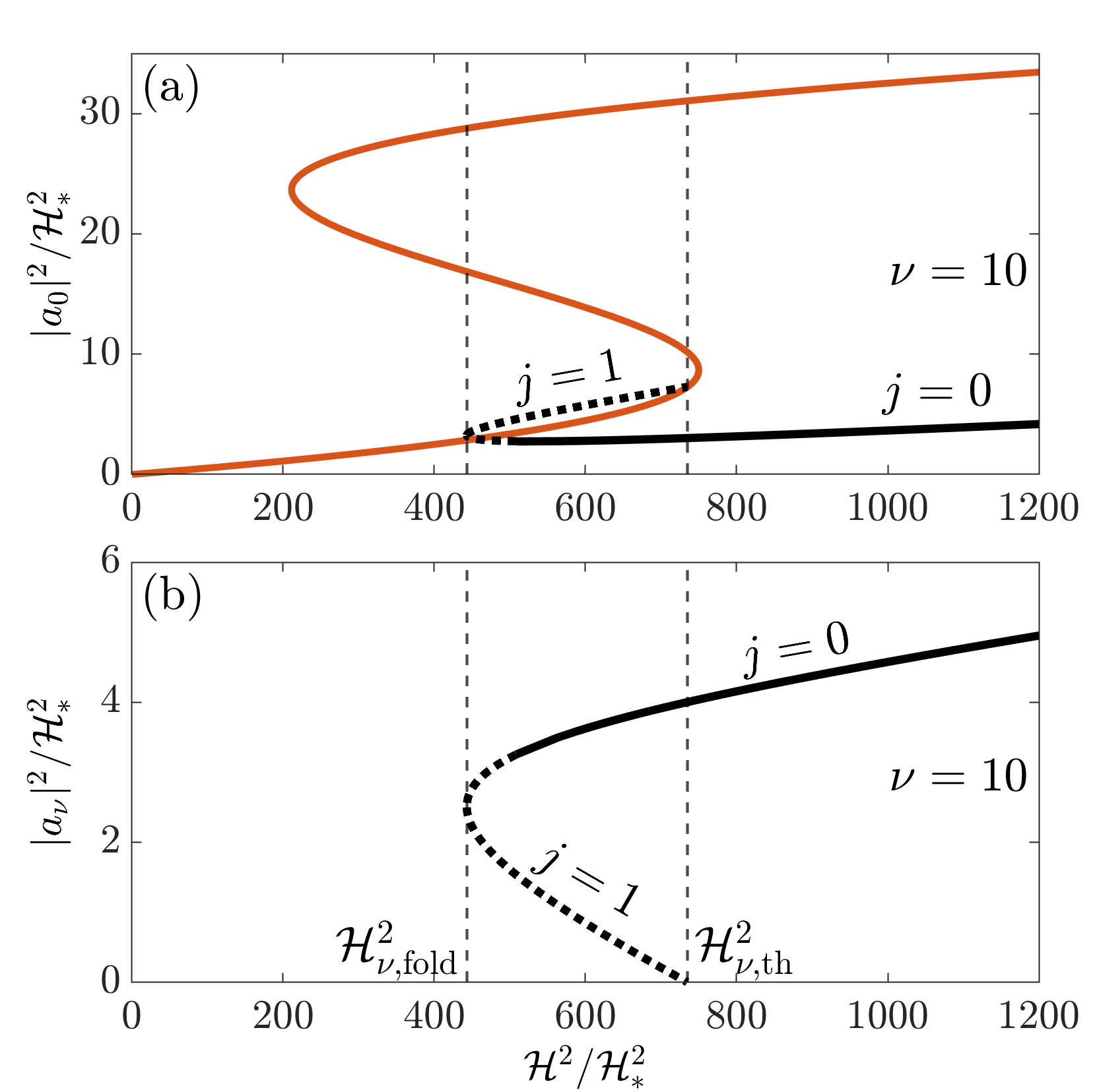}	     	
	\caption{Sub-critical bifurcation of the  OPO  state, when the pump power is used as a control parameter: $\nu=10$, $\delta_{0a}/\kappa_a = 7$, $\ep=0$. (a)  The  black line shows the zero-mode power, $|\hat a_0|^2$,  for the OPO state, see Eq.~\bref{eq8}. The red (gray) line shows the power of the homogeneous state, $|\wt a_0|^2$.	(b)~The OPO sideband power, $|\hat a_{10}|^2$, vs pump. The full and dotted parts of the black lines in (a) and (b) correspond to $j=0$ and $j=1$ in Eq.~\bref{x3a}. 
	}
	\lab{f3}
\end{figure}

\begin{figure}[t]
	\centering
	\includegraphics[width=0.49\textwidth]{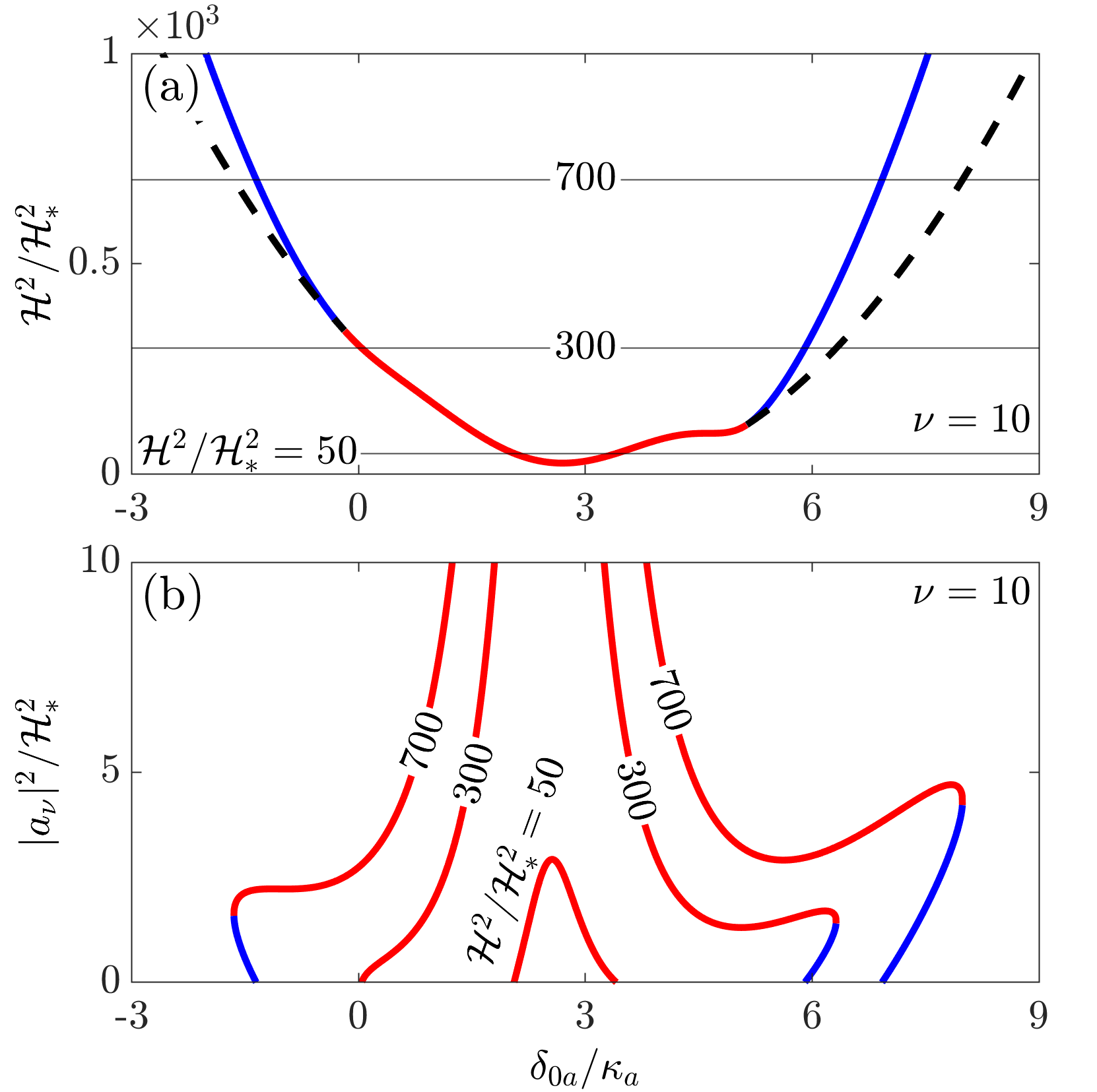}	     
	\caption{(a) The range of existence of the $\nu=10$ OPO state in the pump power-detuning parameter space. The red/blue (gray/dark gray)line corresponds to the supercritical/subcritical bifurcation. The dashed line is the fold condition, see Eq.~\bref{g1} and Fig.~\ref{f3}(b), where the two OPO states merge.  (b)  Powers of the OPO sidebands vs detuning for three selected pump powers: $\cH^2/\cH_*^2=50$ ($\cW\approx 0.7\mu$W), $\cH^2/\cH_*^2=300$ ($\cW\approx 4\mu$W), $\cH^2/\cH_*^2=700$ ($\cW\approx 9\mu$W). 
	}
	\lab{f4}
\end{figure}

Computer algebra helps to demonstrate 
that Eqs.~\bref{qe} and \bref{x3a} reduce to a quadratic equation for $|\hat a_0|^2$, 
\be
Q_{2} \frac{|\hat a_0|^{4}}{\mathcal{H}_*^{4}} 
+ Q_{1} \frac{|\hat a_0|^{2}}{\mathcal{H}_*^{2}} 
+ Q_{0}= 0,
\lab{eq8}
\ee
where
\be
\begin{split}
	Q_{2} &=   \left(|\vD_{0 a}|^2 + |\vD_{\nu a}|^2 \right)^2 - 4 \text{Re}^2 \left(\vD_{0 a} \vD_{\nu a}^* \right),\\ 
	Q_{1} &= \left(|\vD_{0 a}|^2 + |\vD_{\nu a}|^2 \right)\\ 
	&\times \Big(4\text{Im} \left(\vD_{0 b} \vD_{\nu a} \right) \text{Im} \left(\vD_{0 a} \vD_{\nu a}^* \right) - 
	\frac{\cH^2}{2\cH_*^2}\Big), \\ 
	Q_{0} &=   \text{Im} \left(\vD_{0 b} \vD_{\nu a} \right) 
	\Big( 4|\vD_{0 a} \vD_{\nu a}|^2 \text{Im} \left(\vD_{0 b} \vD_{\nu a} \right)\\ 
	&-  \text{Im} \left(\vD_{0 a} \vD_{\nu a}^*\right) \frac{\cH^2}{2\cH_*^2} \Big) + \frac{\cH^4}{16\cH_*^4}.
\end{split}
\lab{e8a}
\ee
A pair of solutions of Eq.~\bref{eq8} degenerate when the discriminant 
becomes zero, i.e., $Q_1^2=4Q_0Q_2$, 
which opens up as 
\be
\begin{split}
		\frac{\mathcal{H}^4_{\nu,\text{fold}}}{\mathcal{H}_*^4} &- 16 \text{Im} \left(\vD_{0 b} \vD_{\nu a} \right) \text{Im} \left(\vD_{0 a} \vD_{\nu a}^* \right) \frac{\mathcal{H}^2_{\nu,\text{fold}}}{\mathcal{H}_*^2} \\ &- 16 \left(|\vD_{0 a}|^2 - |\vD_{\nu a}|^2 \right)^2 \text{Im}^2 \left(\vD_{0 b} \vD_{\nu a} \right) = 0.
\end{split}
\lab{g1}
\ee
Eq.~\bref{g1} corresponds to the fold line in the parameter space.

Diagrams in Figs.~\ref{f1}(d)-\ref{f1}(e) are plotted for the intraresonator mode power varying with detuning while the laser power is kept fixed. Expressing thresholds in terms of the laser power is also possible. Substituting Eq.~\bref{e77} in Eq.~\bref{cub} and using Eq.~\bref{eq7}, provides the threshold values of the laser power,
\be
\begin{split}
	\frac{\cH_{\nu,\text{th}}^2}{\cH_*^2}&=
	4|\vD_{0 b} \vD_{\nu a}| \left(|\vD_{0 a}|^2 +  |\vD_{\nu a}|^2\right)\\ &-
	8\text{Re} \left(\vD_{0 b}\vD_{0 a}  \right) |\vD_{\nu a}|^2.
\end{split}
\lab{eq9}
\ee
The fold line, $\cH^2=\cH_{\nu,\text{fold}}^2$, splits from the threshold condition, $\cH^2=\cH_{\nu,\text{th}}^2$ at the points where the bifurcation changes from super- 
to sub-critical, see Fig.~\ref{f4}(a). 

The detuning dependencies of the sideband amplitudes in Fig.~\ref{f4}(b) show in an obvious way how the supercritical bifurcations change to sub-critical as the pump power is increased. Fig.~\ref{f2}(c) compares the analytical sideband amplitudes (red lines)
 with the ones found from the numerical modelling of the master equation. Fig.~\ref{f5} extends this comparison to the higher powers, where the OPO bifurcations become sub-critical.

\begin{figure}[t]
	\centering
	\includegraphics[width=0.5\textwidth]{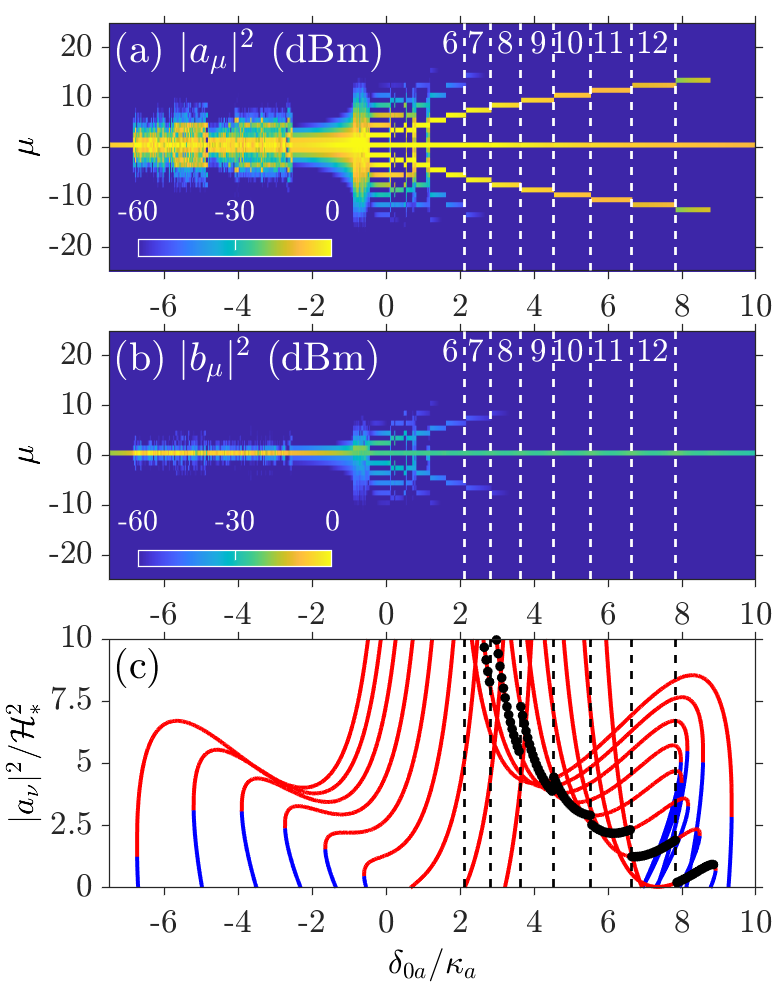}	     
	\caption{(a,b) Results of dynamical simulations of Eq.~\bref{tp1}. Panel (a) shows 
		the mode number spectrum for the pump and panel (b) for the second harmonic  field, $\cH^2/\cH_*^2=700$ ($\cW\approx 9\mu$W), $\ep=0$. (c) Side-band amplitudes of the pump field found using Eq.~\bref{x3a} (red (gray) and blue (dar gray) lines) and from the data in the panel (a) (black circles). }
	\lab{f5}
\end{figure}

\section{Eckhaus instabilities}
 Figures~\ref{f2} and \ref{f5} show that as detuning approaches zero and crosses to the positive values, the resonator enters a regime of operation characterised by the ladder-like transitions between the sequential OPO states, see Eq.~\bref{op}.   Points where one state switches to the next correspond to the Eckhaus instabilities, i.e., to the instabilities of the 
 $\hat a_{0,\nu}+a_{\nu}e^{i\nu\ta}+a_{-\nu}e^{-i\nu\ta}$ pattern relative to
$\hat a_{0,\nu+1}+a_{\nu+1}e^{i(\nu+1)\ta}+a_{-(\nu+1)}e^{-i(\nu+1)\ta}$~\cite{comphys}. Here, we have added the subscript $\nu$, $\hat a_{0}\to \hat a_{0,\nu}$, $\hat b_{0}\to \hat b_{0,\nu}$, to indicate explicitly that the amplitude of the zero mode in an OPO state depends on the order $\nu$ of the sidebands involved; see Eqs.~\bref{eq8} and \bref{e8a} for $\hat a_{0}$, and  Eq.~\bref{t2b} for $\hat b_{0}$. Refs.~\cite{longhi,per} provide discussions and comprehensive overviews of the Eckhaus instabilities in optical systems using an assumption of the unbounded geometry, while Ref.~\cite{comphys} and the present manuscript use the coupled-mode theory capturing the finite-size effects important in microresonators.

We now consider the $\pm\mu$ sideband pair as a small perturbation disturbing the OPO state of order $\nu$. $\nu\ne 0$ corresponds to an OPO state with  $b_0=\hat b_{0,\nu}$ and $\nu=0$ to the homogeneous state $b_0=\wt b_0$. 
 An equation for the instability growth rate is then easily derived from Eqs.~\bref{p92} and \bref{p93} applying the substitution $a_\mu(t)=x_{\mu a} e^{t\lambda_{\nu,\mu}}$ and $a_{-\mu}^*(t)=x_{-\mu a} e^{t\lambda_{\nu,\mu}}$,
\be
\lambda_{\nu,\mu}=-\tfrac{1}{2}\kappa_a+\sqrt{\gamma_a^2|b_0|^2-\kappa_a^2|\vD_{\mu a}|^2},~\mu\ne 0.
\lab{in1}
\ee
For $\nu=0$ (Benjamin-Feir instability), we shall apply  $|b_0|^2=|\wt b_0|^2=\gamma_b^2|\wt a_0|^4/\kappa_b^2|\vD_{0b}|^2$. Here, $|\wt a_0|^2$ is a function 
of the laser power, $\cH^2$, see Eq.~\bref{qn}. For
$\nu\ne 0$ (Eckhaus instability),  $|b_0|^2=|\hat b_{0,\nu}|^2=\kappa_a^2|\vD_{\nu a}|^2/\gamma_a^2$, which does not  explicitly involve $\cH^2$. We should, however, stress that the range of existence of $|a_\nu|^2>0$ and, hence, of  $\hat b_{0,\nu}$ (via its phase) is power dependent, see Eq.~\bref{t33} and Fig.~\ref{f4}(b). 

The growth rate of the Eckhaus instabilities, i.e.,  $\nu\ne 0$, of the OPO states simplifies to 
\be
\begin{split}
&\lambda_{\nu,\mu}=-\tfrac{1}{2}\kappa_a+\kappa_a\sqrt{|\vD_{\nu a}|^2-|\vD_{\mu a}|^2}=\\
&-\tfrac{1}{2}\kappa_a+\sqrt{(\delta_{0a}+\tfrac{1}{2}D_{2a}\nu^2)^2-
	(\delta_{0 a}+\tfrac{1}{2}D_{2a}\mu^2)^2}.
\end{split}
\lab{in2}
\ee
The Eckhaus instability rates for the $\nu=5$ and $\nu=6$ OPO states vs detuning are shown in Figs.~\ref{f2}(d) and \ref{f2}(e), respectively. The intervals of stability match the results of dynamical simulations; see the vertical dashed lines in Figs.~\ref{f2} and \ref{f5} which are derived from the conditions Re$\lambda_{\nu,\nu\pm 1}=0$. 

Thus, we have demonstrated that the OPO tuning from one mode pair to the other is underpinned by a sequence of Eckhaus instabilities developing for positive detunings. Changing dispersion at the pump frequency to anomalous, $D_{2a}>0$, moves the Eckhaus range to the negative detunings.

\begin{figure*}[t]
 	\centering
 	\includegraphics[width=1.\textwidth]{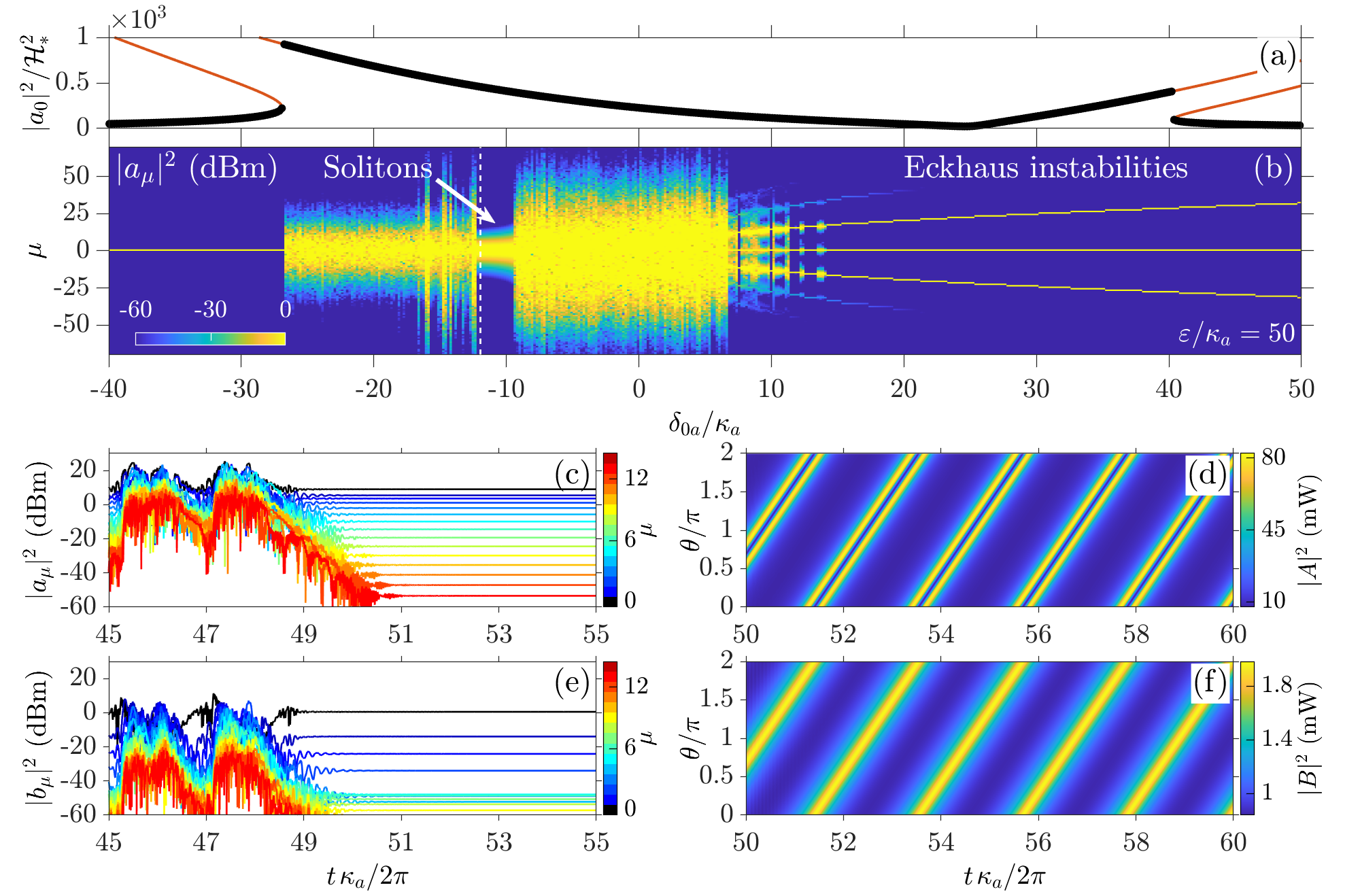}	     
 	\caption{Detuning scan data (a,b) and the soliton modelocking data (c-f) for $\ep/\kappa_a=50$,  $\cH^2/\cH_*^2=310000$ ($\cW=4$mW), see Eq.~\bref{tp1}. (a) shows the bistability of the homogeneous state. Black points indicate solutions used as initial conditions for data in (b).
 	(c,e) transition to modelocking for the detuning marked by the dashed white line in (b). The colour bars show the mode numbers, $\mu=0,1,\dots 16$. (d,f) show the space-time evolution of the soliton in the reference frame rotating with the rate $D_{1a}$. The colour bars show the intraresonator power for the pump (d) and second harmonic (f).}
 	\lab{f7}
\end{figure*}

\begin{figure*}[t]
	\centering
	\includegraphics[width=1.\textwidth]{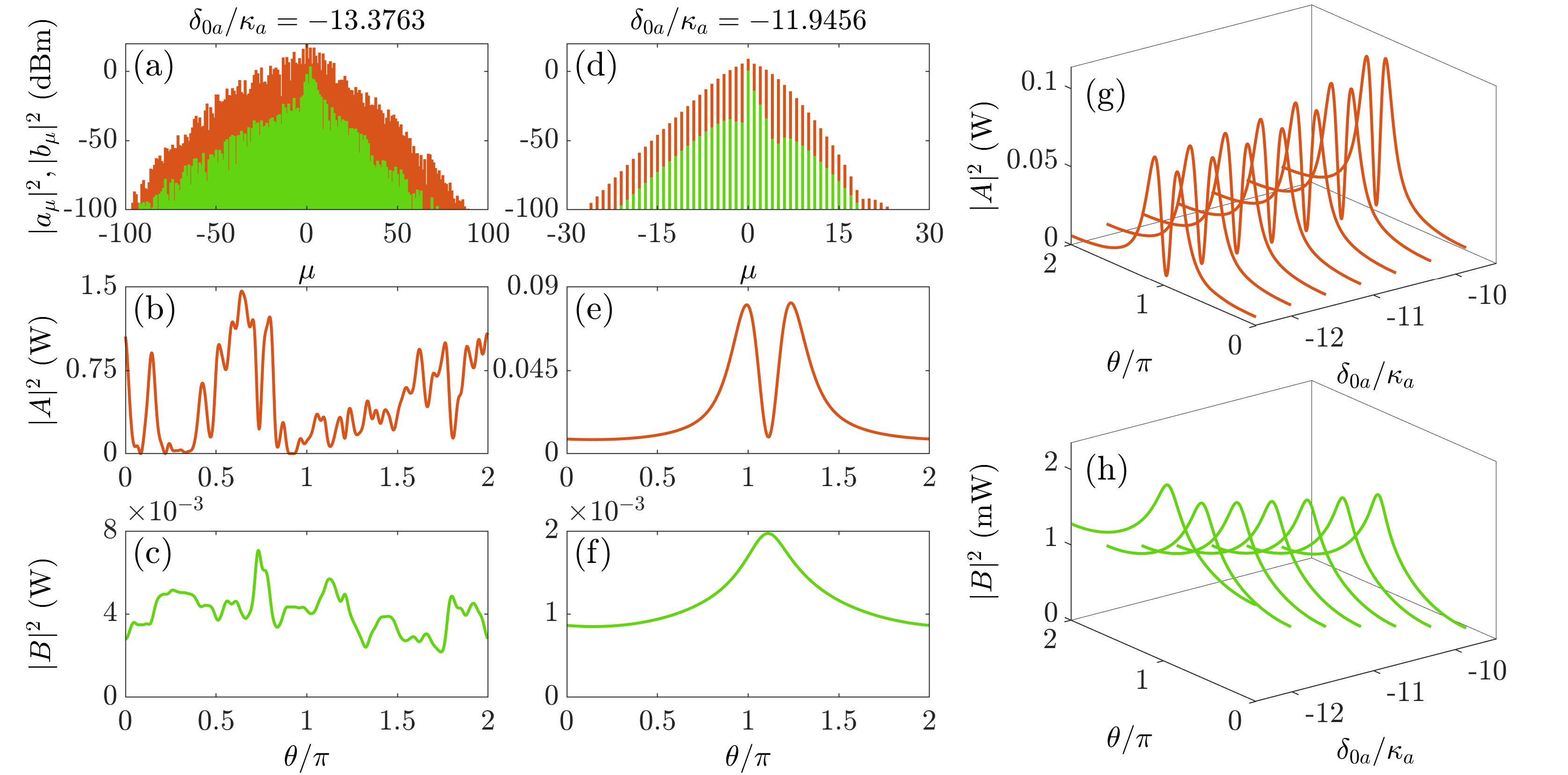}	  	     
	\caption{Details of the spectra and spatial profiles of the pump (red (gray)) and second harmonic (green (light gray)) outside  (a-c) and inside the soliton modelocking interval (d-f). (g,h) show how the spatial soliton profile varies with detuning. The system parameters are as in Fig.~\ref{f7}.}
	\lab{f8}
\end{figure*}

\begin{figure*}[t]
	\centering
	\includegraphics[width=1.\textwidth]{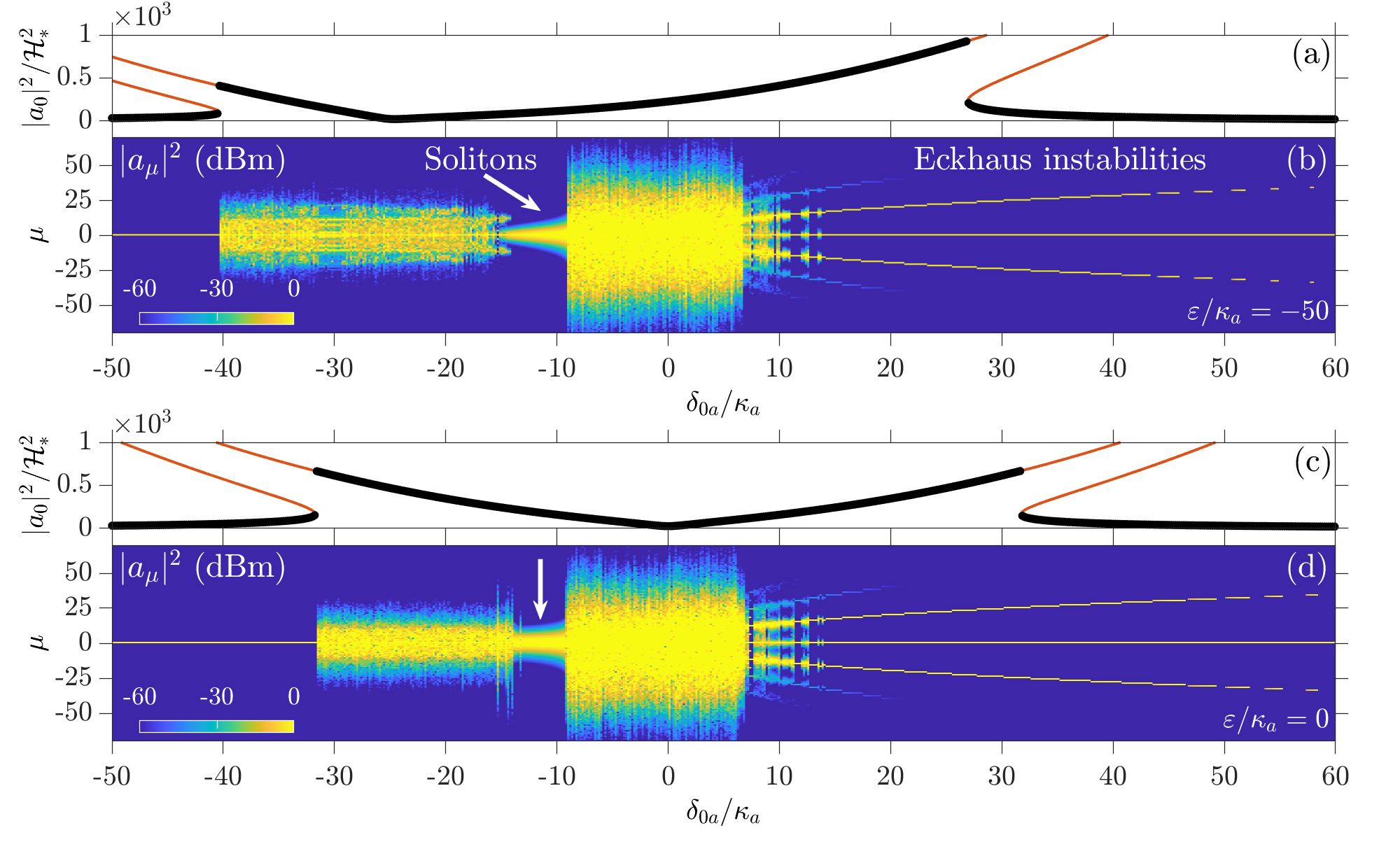}	  	     
	\caption{Detuning scan data  for $\ep/\kappa_a=-50$ (a,b) and $\ep/\kappa_a=0$ (c,d):  $\cH^2/\cH_*^2=310000$ ($\cW=4$mW), see Eq.~\bref{tp1}. (a,c) show the bistability of the homogeneous state. Black points indicate solutions used as initial conditions for data in (b,d).}
	\lab{f9}
\end{figure*}

\section{Two-hump solitons}
Spectra in Fig.~\ref{f5}(a) correspond to the chaotic dynamics if $-7\lessapprox\delta_{0a}/\kappa_a\lessapprox -3$, while for
$-3\lessapprox\delta_{0a}/\kappa_a\lessapprox 0$, the resonator 
enters modelocking. Further systematic numerical exploration 
of the modelocking has revealed that the higher powers 
bring more modes to the phase-locked state, which corresponds 
to a train of the soliton pulses.

Figures~\ref{f7}-\ref{f9} show the data sets we have generated solving the master model, Eq.~\bref{tp1}, for the laser power $\cW=4$mW and  three different values of the phase mismatch parameter,
$\ep/\kappa_a=+50$ (Figs.~\ref{f7} and \ref{f8}), $\ep/\kappa_a=-50$ (Fig.~\ref{f9}(b)), and $\ep/\kappa_a=0$  (Fig.~\ref{f9}(d)).
For all three cases, we have found the range of the negative detunings located between the two tilted resonances of the homogeneous state, where the chaotic multimode dynamics cease to exist and is replaced by modelocking and soliton formation. The numerical simulations have been initialised by the noise on top of the homogeneous states; see the black crosses along the red lines in Figs.~\ref{f7}(a),
\ref{f9}(a) and \ref{f9}(c).   For positive detuning, all homogeneous states are unstable, and independently from the initial condition, the system converges to the sequence of the Eckhaus instabilities. For negative detuning, the low power homogeneous state is mostly stable, and therefore the resonator transits to chaos after the bistability range ends.

Dynamically, the transition from chaos to modelocking happens
after sufficiently long simulation time that have typically been extended to one or  few hundreds of photon lifetimes, $2\pi/\kappa_a$. The emergence of modelocking for the modes $\mu=0,1,2,\dots, 16$ is shown in Figs.~\ref{f7}(c) and \ref{f7}(e). The space-time shapes of the modelocked soliton pulses are shown in Figs.~\ref{f7}(d), \ref{f7}(f) and Figs.~\ref{f8}(e)-\ref{f8}(h).   The soliton in the pump field has two humps, while the single pulse in the second harmonic is centred between them. 
The solitons can be compared with the chaotic waveforms shown in  Figs.~\ref{f8}(a)-\ref{f8}(c).

Spectral shape of the pump component of the soliton is a typical soliton spectrum, which gradually decays as $\mu$ moves away from the centre; see Figs.~\ref{f7}(c) and \ref{f8}(d). The spectrum of the second harmonic is, however, different. Its only significant modes are $\mu=0,1,2$ and, perhaps, $3$; see Figs.~\ref{f7}(e) and \ref{f8}(d).
It prompts a hypothesis that the soliton could be qualitatively considered as a broadband pump pulse supported by the effective potential created by the few dominant modes in the second harmonic field. These modes are near-phase-matched
and, therefore, are subjected to the optical Pockels effect~\cite{pra,pock}. 
On the contrary, the solitons' spectral tails are phase-mismatched via growing $\mu|D_{1a}-D_{ab}|$ and, therefore, experience the effective (cascaded) Kerr nonlinearity~\cite{pra,comphys}. Pronounced spectral reshaping around $\mu=0$, see Fig.~\ref{f8}(d), signals the transition between the Pockels and Kerr nonlinearities. The spectral tails are generated via the frequency-sum and frequency-difference terms entering the sums in the master model, see Eq.~\bref{tp1}, and neglected in the approximate theory describing the OPO regime, see Eq.~\bref{cm}.

Solitons reported here have further features different from what is commonly known for solitons in optical resonators. First, these solitons are shifted well outside the bistability of the homogeneous state and are located between the two bistability intervals. Increasing $\ep$, and depending on its sign, starts destroying the bistability for either negative or positive detunings, see Fig.~\ref{f2}, and shrinks the range where solitons are observed. For $|\ep|$ becoming large, $|\ep|\sim\mu|D_{1a}-D_{1b}|$,  the other type of solitons is emerging~Ref.~\cite{pra}. The solitons in Ref.~\cite{pra} are more conventional in the sense that they co-exist with the bistability and are supported by the familiar interplay between dispersion and the effective Kerr effect, albeit derived via the dressed-state theory. 

Another interesting property of the soliton profiles in Fig.~\ref{f8} is that the finite size effects responsible for the spectral discreteness and, therefore, making a difference between the frequency-comb solitons and the solitons with continuous spectra, play a role here. In particular, one can see the soliton background is slightly curved. We have checked that changing the dispersion sign to anomalous moves solitons to the positive detunings and keeps them outside the bistability range. If the two-hump solitons found here connect or not to the solitons reported in Ref.~\cite{pra} and understanding of the full range of the soliton existence are the problems left for future analysis.

\section{Summary}
We have presented a theory of parametric conversion via the second-harmonic generation in the phase-matched whispering gallery LiNbO$_3$ microresonators with large walk-off. The pump wavelength was assumed at one micron, so the dispersion was considered normal. The large walk-off condition, which means that the repetition rate difference is much larger than the phase-matching parameter, $\mu |D_{1a}-D_{1b}|\gg \ep$, applied in this work complements our recent study of the resonators with $\mu |D_{1a}-D_{1b}|\sim \ep$, see Ref.~\cite{pra}. Here, $\mu$ is the relative mode number counted from the pump.

We have demonstrated that tuning of the parametric signal and generation of sideband pairs around the pump is associated with the sequence of the Eckhaus instabilities happening for positive detunings. We have derived a transparent approximate expression for the growth rates of the Eckhaus instabilities, see Eq.~\bref{in2} and Fig.~\ref{f2}. A feature of these instabilities is their quasi-independence from the walk-off due to the second harmonic staying quasi-monochromatic and most of the spectrum generated in the pump field. We also found the sideband powers and identified conditions for the super- and sub-critical transitions to the OPO, i.e.,  Turing-pattern, states. All these results are obtained under the conditions of the ring geometry, i.e., when the mode numbers are quantized.

The resonator operation regimes on the negative detuning side are more complex and often chaotic. However, we have found an interval of negative detunings where the multimode chaos is replaced by modelocking. The modelocking leads to the soliton formation in the form of a double-hump pulse in the pump field and the single-hump pulse in the second harmonic. A distinct feature of these solitons is their size comparable to the resonator circumference. They exist on the non-flat background and outside the bistability range of the single-mode state. A spectrum of the pump component of the soliton has a typical near-triangular shape, while only a few first modes dominate the second harmonic component.

Data presented in this study are openly available from
the University of Bath Research Data Archive~\cite{data}.

\begin{acknowledgments}
This work was supported by the European Union Horizon 2020 Framework Programme (812818).
\end{acknowledgments}

\end{document}